\documentclass[aps,twocolumn,preprintnumbers,amsmath,amssymb,superscriptaddress,floatfix,nofootinbib]{revtex4}
\usepackage{graphicx,color,dcolumn,booktabs,bm}
\usepackage{longtable,lscape}
\usepackage{txfonts}
\usepackage{overpic}
\usepackage{epsfig}
\usepackage{amssymb}
\usepackage{rotating}
\usepackage{epstopdf}
\usepackage{appendix}
\usepackage{braket}
\usepackage{indentfirst}
\usepackage{feynmf}   
\usepackage{slashed}  
\usepackage{cases}
\usepackage{color}
\usepackage{multirow}
\usepackage{graphicx,color,dcolumn,booktabs,bm}
\usepackage{cases}
\usepackage{array}
\usepackage{graphicx}
\usepackage{subfigure}

\graphicspath{{Figures/}} %

\usepackage[colorlinks, citecolor=blue,anchorcolor=red,menucolor=red, linkcolor=red,filecolor=red,runcolor=red,urlcolor=blue,frenchlinks=red]{hyperref}

\begin{document}
\title{Production of the $\Xi N$ dibaryon as a weakly bound system in $pp$ collisions}
\author{Tian-Chen Wu} 
\affiliation{School of Physics, Beihang 
 University, Beijing 102206, China}

\author{Atsushi Hosaka} \email{hosaka@rcnp.osaka-u.ac.jp} %

\affiliation{Research Center for Nuclear Physics (RCNP), Ibaraki, Osaka 567-0047, Japan}
\affiliation{Advanced Science Research Center, Japan Atomic Energy Agency, Tokai, Ibaraki 319-1195, Japan}

\author{Li-Sheng Geng}
\email{lisheng.geng@buaa.edu.cn}

\affiliation{School of Physics, Beihang University, Beijing 102206, China}
\affiliation{Peng Huanwu Collaborative Center for Research and Education, Beihang University, Beijing 100191, China}
\affiliation{Beijing Key Laboratory of Advanced Nuclear Materials and Physics, Beihang University, Beijing 102206, China}
\affiliation{Southern Center for Nuclear-Science Theory (SCNT), Institute of Modern Physics, Chinese Academy of Sciences, Huizhou 516000, Guangdong Province, China}

\begin{abstract} 
The $\Xi N$ interaction plays an important role in our understanding on the long-anticipated $H$-dibaryon. 
Recent lattice QCD calculations verified the attractive nature of the $\Xi N$ interaction. On the other hand, whether it is strong enough to generate a bound state remains inconclusive.
In this work, assuming that it can generate a weakly bound state, we study the yields of the $\Xi N$ dibaryon for different binding energies in $pp$ collisions at 7 TeV using the coalescence model and the transport model PACIAE. The yields are estimated first numerically and then analytically adopting a Yukawa-type wave function. In particular, we find that in the weak binding limit, there exists a universal relation between the yield and the binding energy, valid for $pp$ collisions.
\end{abstract}

\maketitle

\section{Introduction}
In the KEK-E373 experiment, the first clear evidence of a deeply bound $\Xi^{-}-^{14}\rm N$ state  was found ~\cite{Nakazawa:2015joa}. However, there is no direct experimental evidence for the existence of the $\Xi N$ dibaryon so far~\cite{ALICE:2019hdt, ALICE:2020mfd, KEK-PSE224:1998trj}. Studying the simplest $\Xi N$ dibaryon (often referred to as the $H$-dibaryon~\cite{Sakai:1999qm})  plays a fundamental role in understanding the non-perturbative strong force in a more complex system, such as $^{14}$N. Recent lattice QCD simulations~\cite{HALQCD:2019wsz} and correlation function studies~\cite{Liu:2022nec} indicate an attractive $\Xi N$ interaction.
With the $\Xi N$ potential from the latest lattice QCD simulation~\cite{HALQCD:2019wsz}, Hiyama et al. found that the $\Xi N$ system cannot bind~\cite{Hiyama:2019kpw}. 
On the other hand, with the ESC08c potential, the $\Xi N$ system can develop a shallow bound state with a binding energy of a few MeV~\cite{Hiyama:2019kpw, Garcilazo:2016gkj, Nagels:2015dia}.
Under such an unsettled situation, in this work we would like to propose an alternative way by studying the production process of the $\Xi N$ dibaryon in $pp$ collisions.

The coalescence model is a well-established method to describe the production process of composite particles. To apply the coalescence model to the $\Xi N$ dibaryon, we need two essential inputs. One is to find a proper process to produce the constituent particles $\Xi$ and $N$. Since the productions of $\Xi$ and $N$ are well studied in LHC experiments, e.g., inelastic proton-proton($pp$) collisions~\cite{ALICE:2010vtz,ALICE:2012yqk, ALICE:2015ial}, 
it is reasonable to choose the $pp$ collisions for this purpose.
In our work, this process is simulated by the transport model PACIAE~\cite{Sa:2011ye}. The other essential part is to properly set the conditions to constrain the constituent particles in phase space. The simplest condition is the cut-off condition~\cite{Scheibl:1998tk, Sombun:2018yqh}, where the constituent particles combine when their relative distance and momentum are smaller than a specified  cut-off. Another microscopic approach is the Wigner density approach~\cite{Gyulassy:1982pe, ExHIC:2010gcb, Scheibl:1998tk, Zhang:2020dwn, Chen:2003tn, Zhang:2020dma, Zhang:2021vsf, Greco:2003mm, Greco:2003xt}, which relies on the wave function of the composite particle. 

If $\Xi$ and $N$ can bind, it is most likely to be a shallow bound state. If the $\Xi N$ dibaryon is a resonance, it is probably a Feshbach resonance, whose seed is a $\Xi N$ bound state coupled with the open channel of $\Lambda \Lambda$. We can make a reasonable assumption that the shape of the wave function of the $\Xi N$ resonance is similar to that of the $\Xi N$ bound state. Therefore, in our work, we assume the $\Xi N$ dibaryon as a weakly bound state. The wave function of the $\Xi N$ dibaryon at large distances should behave as the Yukawa function for a short-range potential. The range of the wave function is only related to the reduced mass and the binding energy $E_B$ of the bound state, 
which means that the yield of $\Xi N$ depends primary on the binding energy $E_B$, which is a universal phenomenon, as we will demonstrate  in this work.

This article is organized as follows. In  Sec.~\ref{formalism}, we briefly introduce the transport model PACIAE and the coalescence model adopting the Yukawa function. The formalism is presented in both numerical and analytic ways. In Sec.~\ref{Results}, we discuss the numerical and analytic results, and then the universal phenomenon. Finally, we present a short summary in Sec.~\ref{Summary}.

\section{Formalism}{\label{formalism}}
The production of a composite particle is naturally divided into two steps: the production of constituent hadrons and their combination into the composite particle. We employ the transport model PACIAE~\cite{Sa:2011ye} for the former step, 
and the coalescence model with the Wigner function approach  for the latter step. 

\subsection{PACIAE model}\label{PACIAE model}
The PACIAE model is a transport model based on the event generator PYTHIA~\cite{Sjostrand:2006za}. It describes high-energy collisions, such as $e^+e^-$ collisions, hadron-hadron collisions, and nucleus-nucleus collisions. Similar to the PYTHIA model,  the PACIAE model also simulates the collision process in terms of parton initiation and hadronization, but with additional transport processes~\cite{Sa:2011ye}. 
 In parton initiation, hadron-hadron collisions are decomposed into parton-parton interactions. The hard part is treated by the leading order perturbative QCD while the soft part involves some phenomenological modelling. Hadronization and decay are then expected after the creation of the mixture of partons. The most significant difference between the transport model PACIAE and the generator PYTHIA is that the former introduces the transport processes, considering the fact the thermodynamic interactions cannot be neglected in the multi-particle states.

The tunable parameters in the PACIAE model are those that determine the probabilities of different quark pairs created from the vacuum, or the parameters of hadronization functions. The PYTHIA Perugia 2011 (P2011) cannot well simulate the experimental yields of $\Xi$ and $\Omega$, where the yield of $\Omega$ is several times smaller than the experimental measurement~\cite{ALICE:2012yqk}. This shows the insufficiency of this simulation mode in describing the production of strange quarks. Moreover, the structure of Nambu-Goldstone bosons is more complicated than the one encoded in the Lund string model. Since in our simulation, we are more interested in the productions of baryons, especially multi-strange baryons, we tuned the parameters so that the simulation results are in better agreement with the experimental yields of $N$, $\Xi$ and $\Omega$, and ignore the discrepancy in the yields of mesons.
\subsection{Coalescence model}\label{Coalescence model}
The basic idea of the coalescence model is that the constituent particles of a shallow-bound composite particle, whose binding energy is small compared to the evolution temperature, are difficult to combine until the whole system reaches the kinetic freeze-out. This implies the final state approximation~\cite{Gyulassy:1982pe}, where  ``final state" indicates that the constituent particles experience almost no interaction with other hadrons. The contribution from intermediate interactions before kinetic freeze-out to the yield of the composite particle can be neglected. 
In addition, the coalescence time is short compared to the interaction time in the ``final state". Therefore, the coalescence process can be described in a sudden approximation~\cite{Scheibl:1998tk}. As a result, we can interpret the formation of composite particles as a trace over the density of the source $\hat{\rho}_S$ in the ``final state", which is the phase space distribution of constituent particles, and the density of the composite particle $\hat{\rho}_C=\ket{\Psi_C}\bra{\Psi_C}$, i.e., $\rm{tr}[\hat{\rho}_S\hat{\rho}_C]=tr[\hat{\rho}_S\ket{\Psi_C}\bra{\Psi_C}]$.

It is important to note that the density of the source is described by a semi-classical transport model while the wave function of the composite particle is from quantum theory. Thus, transformation is needed to combine these two models. For this, the Wigner transform is an effective method. In this approach, both the density of the source $\hat{\rho}_S$ and the density of the composite particle $\hat{\rho}_C$ are transformed to Wigner densities $\hat{\rho}_S^W$ and $\hat{\rho}_C^W$.

The $n$-body Wigner density of the source calculated from the transport model can be written as~\cite{Scheibl:1998tk}
\begin{equation}
\begin{aligned}\label{Eq:source density}
\hat{\rho}_S^W(\bm x_1,\bm p_1,\cdots,\bm x_{n}, \bm p_{n})= \left \langle \sum_{(c)} \prod_{i=1}^{n} (2\pi)^3 \delta^3(\bm{x}_i-\bm{\Tilde{x}}_i) \delta^3(\bm{p}_i-\bm{\Tilde{p}}_i)  \right \rangle,
\end{aligned}
\end{equation}
where $(c)$ is the index of combinations in a collision event, $i$ is the index of hadrons in each combination, $\bm{\Tilde{x}}_i$ and $\bm{\Tilde{p}}_i$ are the phase space coordinates of particle $i$ in the PACIAE final state, and $\left\langle\cdots\right\rangle$ denotes that the result is averaged over all the event runs. The Wigner density of the composite particle can be obtained by the Wigner transform
\begin{widetext}
\begin{equation}
\begin{aligned}\label{Eq:cluster density}
\hat{\rho}_C^W(\bm r_1,\bm q_1,\cdots,\bm r_{n-1}, \bm q_{n-1})=&\int\Psi_C(\bm{r}_1+\frac{1}{2}\bm{y}_1,\cdots,\bm{r}_{n-1}+\frac{1}{2}\bm{y}_{n-1})\Psi_C^{\ast}(\bm{r}_1-\frac{1}{2}\bm{y}_1,\cdots,\bm{r}_{n-1}-\frac{1}{2}\bm{y}_{n-1})\\&
 e^{-i\bm{q}_1 \cdot \bm{y}_1}\cdots e^{-i\bm{q}_{n-1} \cdot \bm{y}_{n-1}} d^3\bm y_1\cdots  d^3\bm y_{n-1},
\end{aligned}
\end{equation}
\end{widetext}
where $\bm r_{n-1}$ and $\bm q_{n-1}$ are the $n-1$ relative coordinates calculated from position and momentum coordinates $\bm x_{n}$ and $\bm p_{n}$. Then the differential and total yield $Y$ of an $n$-body system averaged in each event is~\cite{Mattiello:1996gq,Scheibl:1998tk}
\begin{widetext}
\begin{equation}
    \begin{aligned}\label{yield}
 &\frac{dY}{d \bm P}= g \int \hat{\rho}_S^W(\bm x_1,\bm p_1,\cdots,\bm x_n, \bm p_n) \hat{\rho}_C^W(\bm r_1,\bm q_1,\cdots,\bm r_{n-1}, \bm q_{n-1}) \delta^3(\bm P-(\bm p_1+\cdots \bm p_n))\frac{d\bm x_{1} d\bm p_{1} }{(2\pi)^3 } \cdots \frac{d\bm x_{n} d\bm p_{n} }{(2\pi)^3 },  \\
& Y= g \left \langle \sum_{(\rm c)} \hat{\rho}_C^W(\Tilde{\bm r}_1,\Tilde{\bm q}_1,\cdots,\Tilde{\bm r}_{n-1}, \Tilde{\bm q}_{n-1}) \right \rangle,
    \end{aligned}
\end{equation}
\end{widetext}
where $\bm P$ is the total momentum of the composite particle, $\Tilde{\bm r}_n$ and $\Tilde{\bm q}_n$ are the relative position and momentum coordinates calculated with the position coordinate $\tilde{\bm x}_n$ and momentum coordinate $\tilde{\bm p}_n$ of the primary hadrons in each combination (c) from the transport model. The additional factor $g$ is the spin statistical factor, which is 1/4 in our work.
\subsection{Use of the Yukawa function}
The potential takes a general Yukawa form for the $\Xi N$ dibaryon\cite{HALQCD:2019wsz}, which vanishes sufficiently fast at large distances. Therefore, the wave function can be approximated to be the Yukawa function at large distances. However, the Yukawa function has a singular point at the origin. This can be avoided by introducing a form factor $\Lambda^2/(q^2+\Lambda^2)$, which characterizes the size of hadrons. We set the cutoff $\Lambda$ at 0.8 GeV corresponding to a size of $\langle r^2 \rangle ^{1/2} \sim 0.6$ fm.  The wave function then has the following form:
\begin{equation}\label{Eq:wavefunction}
\Psi(r)=A\left(\frac{e^{-\beta r}}{r}-\frac{e^{-\Lambda r}}{r}\right),
\end{equation}
where $\beta=\sqrt{2 \mu E_B}$, $\mu$ is the reduced mass,  $E_B$ is the binding energy $E_B$, and $A$ is the normalization constant $A=\sqrt{\beta \Lambda (\beta+\Lambda)/(2\pi(\beta-\Lambda)^2)}$. 

Since the analytical form of the Wigner density corresponding to the wave function in Eq. (\ref{Eq:wavefunction}) is hard to obtain, we expand the wave function in terms of the Gaussian bases,
\begin{equation}
\begin{aligned}\label{Eq:gaussian}
\Psi(r)= \sum_{i=1}^N c_i\left(\frac{2\omega_i}{\pi}\right)^{3/4} e^{-\omega_i r^2},
\end{aligned}
\end{equation}
where $N=50$ is the number of bases, $\omega_i$ characterizes the width of the Gaussian bases, and $c_i$ is the corresponding weighting factor.
With this expansion, the Wigner density has an analytical form,
\begin{equation}
\begin{aligned}\label{Eq:gaussian Wigner density}
\hat{\rho}_C^W (\bm{r},\bm{q})=&8\sum_{i=1}^N c_i^2 \exp \left( -2\omega_i r^2 - \frac{q^2}{2\omega_i}\right )\\
&+16\sum_{i>j}^N c_i c_j \left(\frac{4\omega_i\omega_j}{(\omega_i+\omega_j)^2}\right)^{3/4} \exp\left(-\frac{4\omega_i\omega_j}{\omega_i+\omega_j} r^2\right)\\
&\times \exp\left(-\frac{q^2}{\omega_i+\omega_j}\right) \cos\left({2\frac{\omega_i-\omega_j}{\omega_i+\omega_j} \bm{r}\cdot \bm{q}}\right).
\end{aligned}
\end{equation}

It is convenient to calculate the yield of $\Xi N$ numerically, but it is not transparent to show the connection between the yield and the binding energy $E_B$. To better understand the relation, we need to make some approximations. According to  the mean value theorem, $\rho_C^W(\Tilde{\bm{r}},\Tilde{\bm{q}})$ for a binding energy $E_B$ in one combination $(c)$ for $\Xi$ and $N$ is
\begin{equation}
\begin{aligned}\label{Eq:analytic Wigner density}
\rho_C^W(\Tilde{\bm{r}},\Tilde{\bm{q}})\sim C_1 A^2 (e^{-\beta C_2}-e^{-\Lambda C_2})\left(\frac{1}{\beta+C_3}-\frac{1}{\Lambda+C_3}\right),
\end{aligned}
\end{equation}
where $C_1$, $C_2$ and $C_3$ are functions of $\Tilde{\bm{r}}$ and $\Tilde{\bm{q}}$.
For the $\Xi N$ dibaryon, the binding energy $E_B$ is small, so $A \sim \sqrt{\beta}$ when $E_B \to 0$. For the same reason, $\frac{1}{\beta+C_3} \sim \frac{1}{C_3}$ and $e^{-\beta C_2}-e^{-\Lambda C_2} \sim e^{-\beta C_2}$. In this weak binding limit, one has
\begin{equation}
\begin{aligned}\label{Eq:analytic Wigner density 2}
\rho_C^W(\Tilde{\bm{r}},\Tilde{\bm{q}})\sim C_1 \beta e^{-\beta C_2}\left(\frac{1}{C_3}-\frac{1}{\Lambda+C_3}\right) \equiv C'\beta e^{-\beta C_2}.
\end{aligned}
\end{equation}
The total yield is the summation of all the combinations. For each combination we have $C'^{(c)}$ and $C_2^{(c)}$, thus the yield $Y$ is
\begin{equation}
\begin{aligned}\label{Eq: approximation 1}
Y=g\left \langle \sum_{(\rm c)} \hat{\rho}_C^W(\Tilde{\bm r},\Tilde{\bm q}) \right \rangle \sim g \beta \left \langle \sum_{(\rm c)} C'^{(c)} e^{-\beta C_2^{(c)}} \right \rangle.
\end{aligned}
\end{equation}
Inspired by the form of Wigner densities of the Yukawa function and converting the sum into an integral, we assume an approximation for Eq. (\ref{Eq: approximation 1}) as follows:
\begin{equation}
\begin{aligned}\label{Eq: approximation 2}
Y \sim g \beta (C_1 e^{-\beta C_2} + P(\beta)) \sim g \sqrt{2 \mu E_B} C_1 e^{-\sqrt{2 \mu E_B} C_2},
\end{aligned}
\end{equation}
where the parameters $C_1$ and $C_2$ can be determined by fitting to the simulation data, and $P(\beta)$ is a polynomial serving as a correction term. In Sec.~\ref{yield of XiN}, the polynomial is taken to be a constant, $P(\beta)=C_3$. In the case of a small binding energy, the yield then has the following asymptotic form
\begin{equation}
\begin{aligned}\label{Eq: approximation 3}
Y \sim \sqrt{E_B} \sim \frac{1}{R},
\end{aligned}
\end{equation}
where $R$ is the root-mean-square radius of the $\Xi N$ dibaryon. In Sec. \ref{yield of XiN}, we will fit this formula  to the simulation data to verify the approximation.
\section{Results and Discussions}{\label{Results}}
\subsection{Productions of normal hadrons}
\begin{figure}[htpb]
\centering
\includegraphics[width=0.45\textwidth]{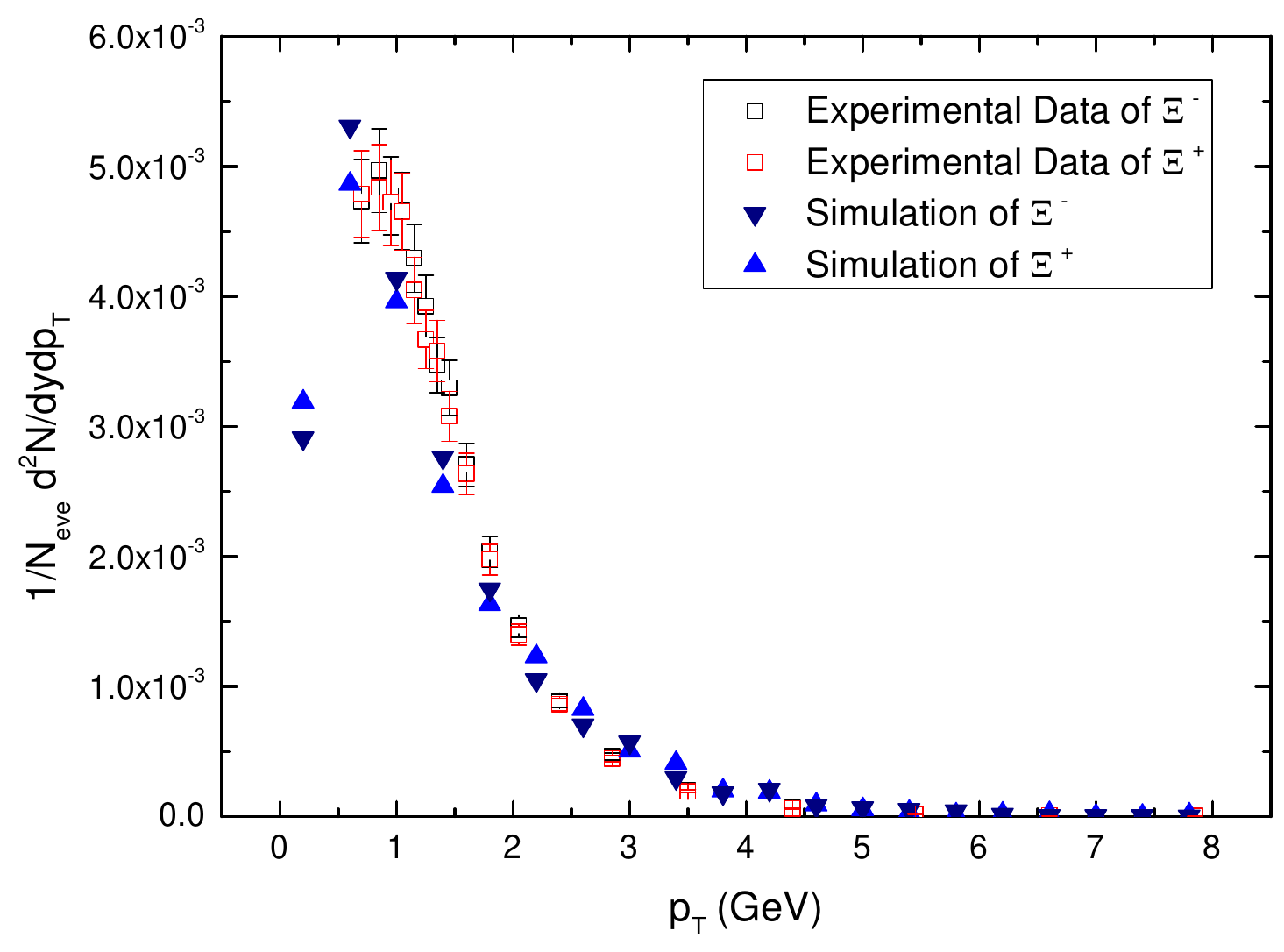}
\caption{ \label{fig: pT distribution 1} $p_T$ distribution of the yield of $\Xi$ in bins of 0.4 GeV, where the squares are the experimental data from ALICE~\cite{ALICE:2012yqk} and the triangles are the simulation results.}
\end{figure}

\begin{figure}[htpb]
\centering
\includegraphics[width=0.45\textwidth]{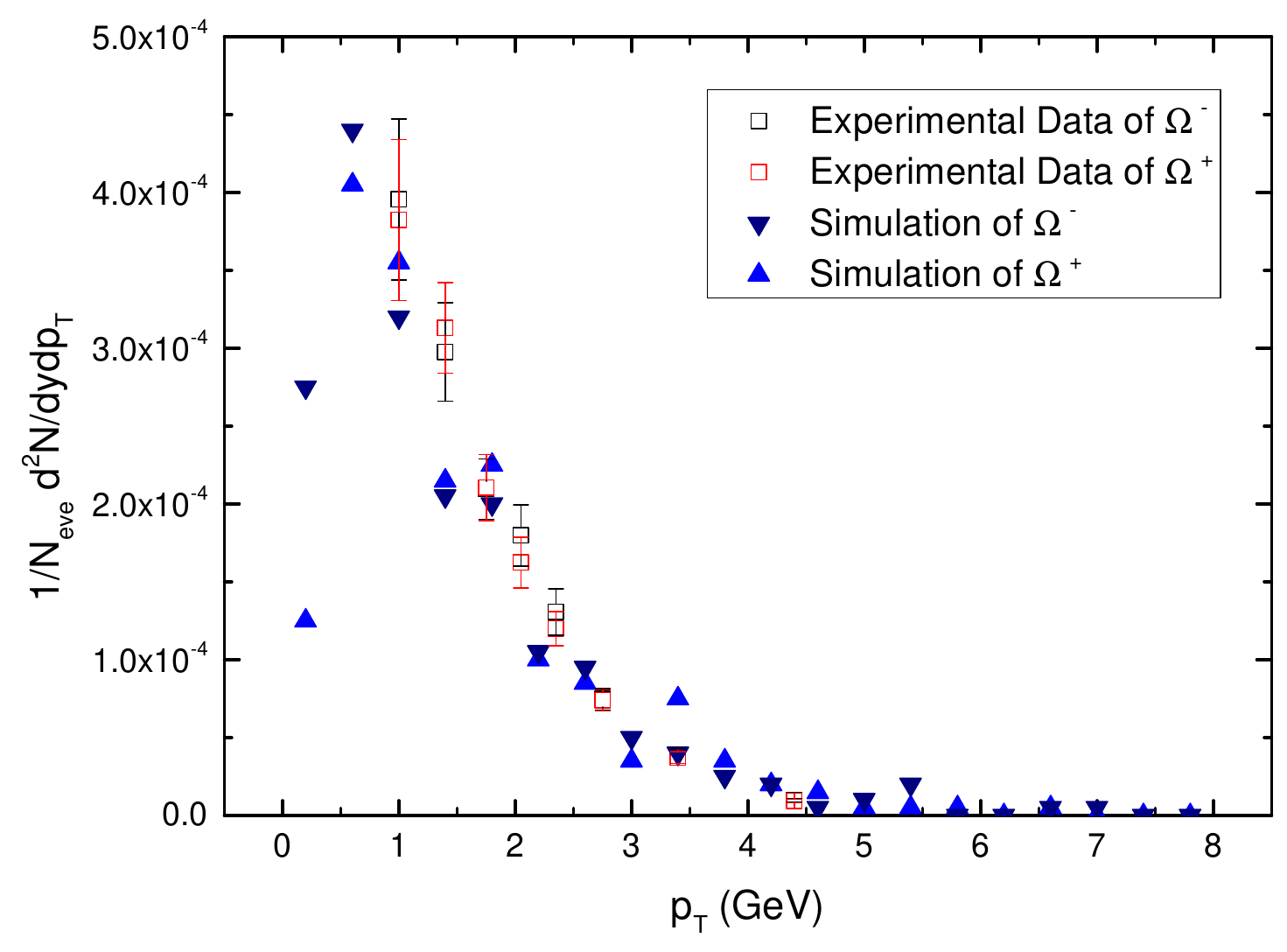}
\caption{ \label{fig: pT distribution 2} $p_T$ distribution of the yield of $\Omega$ in bins of 0.4 GeV, where the squares are the experimental data from ALICE~\cite{ALICE:2012yqk} and the triangles are the simulation results.}
\end{figure}

First, let us obtain the yields of normal hadrons in the PACIAE simulation. As mentioned in Sec.~\ref{PACIAE model}, we focus on the productions of baryons, especially strange baryons, so we have mainly tuned the parameters related to the $s$ quark.
We kept all the parameters at their default values, except for PARJ(1) = 0.06, PARJ(2) = 0.44, and PARJ(3) =  0.8, where PARJ(1) (Default = 0.10) is for the suppression of diquark-antidiquark pair production compared with quark-antiquark production, PARJ(2) (Default = 0.30) for the suppression of $s$ quark pair production with $u$ or $d$ pair production, and PARJ(3) (Default = 0.4) for the extra suppression of strange diquark production compared with the normal suppression of the strange quark~\cite{Sjostrand:2006za}. The so-obtained simulation results for the total yields of baryons are shown in Table \ref{tab: Total yields of baryons}, and the $p_T$ distributions of the yields of $\Xi$ and $\Omega$ are shown in Fig.~\ref{fig: pT distribution 1} and Fig.~\ref{fig: pT distribution 2}. We find that the simulation results are in good agreement with the experimental data (The data are selected in the rapidity region $\left|y\right|<0.5$, and therefore, they are missing in the small transverse momentum region.).
\begin{table*}[htpb]
\caption{\label{tab: Total yields of baryons}Experimental and simulated yields of primary hadrons per $pp$ collision event at $\sqrt{s}$ = 7 TeV.}
\setlength{\tabcolsep}{3.2pt}
\begin{tabular}{cccc}
\hline
\hline
Particle&Data~\cite{ALICE:2012yqk,ALICE:2015ial}&Simulation (Tuned)& Simulation (Default)\\ \hline
 $\Xi^- \ (\times 10^{-3})$ &8.0$\pm 0.1_{-0.5}^{+0.7}$ &7.93&3.78\\
 $\Xi^+ \ (\times 10^{-3})$ &7.8$\pm 0.1_{-0.5}^{+0.7}$&8.06&3.72\\
$\Omega^- \ (\times 10^{-3})$ &0.67$\pm0.03_{-0.07}^{+0.08}$&0.684&0.120\\
$\Omega^+ \ (\times 10^{-3})$ &0.68$\pm0.03_{-0.06}^{+0.08}$&0.728&0.106\\
$p$ &0.124$\pm$0.009 &0.132&0.210\\
$\bar{p}$ &0.123$\pm$0.010 &0.131&0.208\\
\hline
\hline
\end{tabular}
\end{table*}
\subsection{Yield of $\Xi N$ dibaryon}\label{yield of XiN}
\begin{figure}[htpb]
\centering
\includegraphics[width=0.45\textwidth]{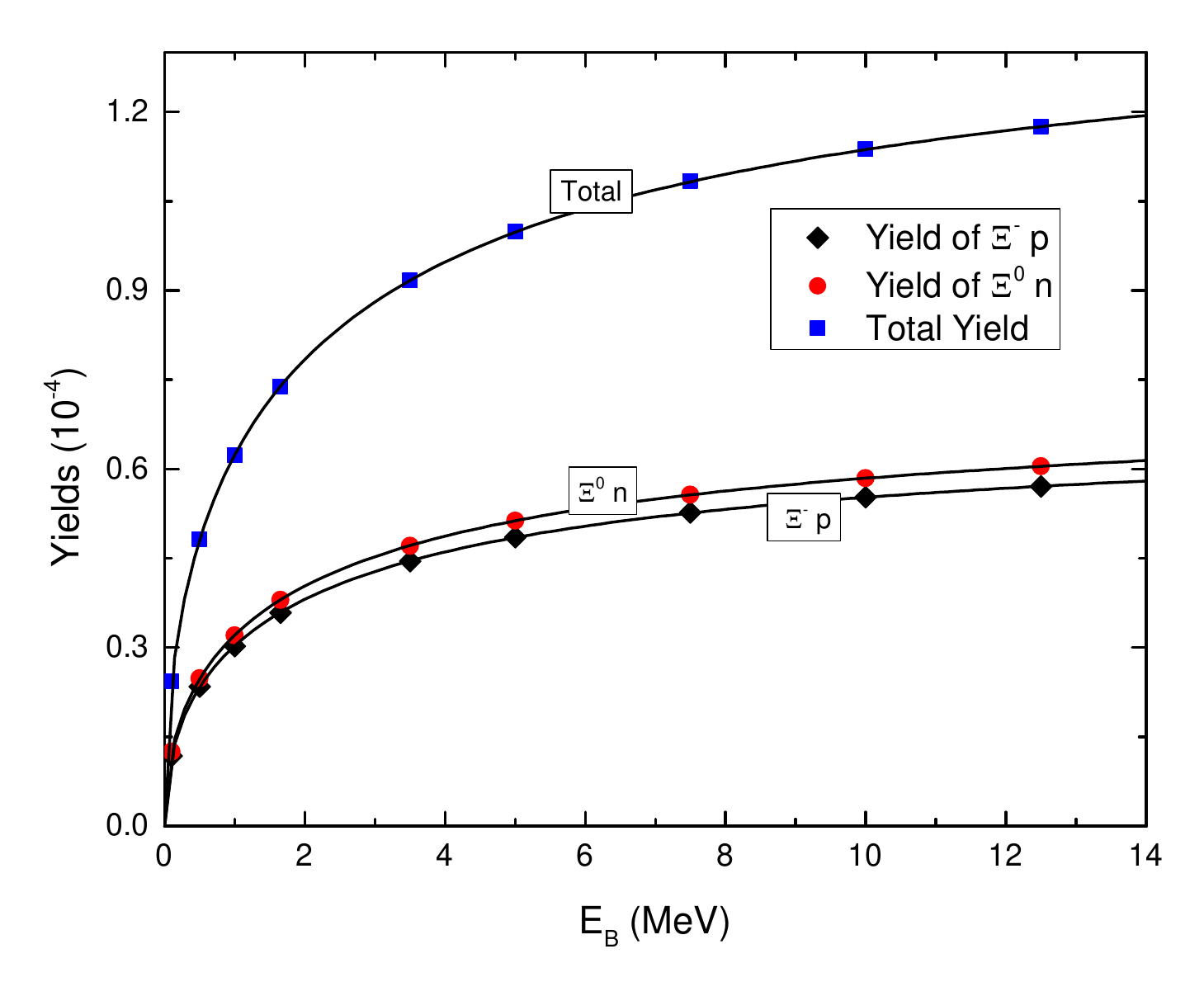}
\caption{ \label{fig: Fitting simulation} Yield of the $\Xi N$ dibaryon for different binding energies $E_B$, where the solid daimonds/squares/circles/ are simulation data, and the lines are the analytic results fitted to the simulation data.}
\end{figure}

With the Wigner density approach, we numerically obtain the yields of the $\Xi N$ dibaryon for different binding energies, which are shown in Table \ref{tab: Total yields of XiN} and Fig. \ref{fig: Fitting simulation}.
\begin{table*}[htpb]
\caption{\label{tab: Total yields of XiN}Averaged yield of $\Xi N$ dibaryon in $pp$ collisions at $\sqrt{s}$ = 7 TeV obtained from the Wigner function approach per $pp$ event, containing charge conjugated states. The binding energy $E_B=1.655$ MeV is that predicted by the ESC08c potential~\cite{Garcilazo:2016gkj}.}
\setlength{\tabcolsep}{3.2pt}
\begin{tabular}{cccccccccc}
\hline
\hline
Binding Energy $E_B$ (MeV)&0.10 & 0.50 & 1.00 & 1.66 & 3.50 & 5.00 & 7.50 & 10.00 & 12.50\\ \hline
$\Xi p \ (\times 10^{-4})$&0.12 & 0.23 & 0.30 & 0.36 & 0.45 & 0.49 & 0.53 & 0.55 & 0.57\\
$\Xi n \ (\times 10^{-4})$&0.13 & 0.25 & 0.32 & 0.38 & 0.47 & 0.51 & 0.56 & 0.59 & 0.60\\
Total $(\times 10^{-4})$& 0.24 & 0.48 & 0.62 & 0.74 & 0.92 & 1.00 & 1.08 & 1.14 & 1.17\\
\hline
\hline
\end{tabular}
\end{table*}
The production yields are of the order of $10^{-4}$, somewhat smaller than those of $\Omega$ by one order of magnitude. However, we expect that the $\Xi N$ dibaryon can be found if it has a binding energy of a few MeV. Although there is no direct experimental evidence for the $\Xi N$ dibaryon so far, we can search for its signature using Femtoscopic techniques~\cite{ALICE:2019hdt, ALICE:2020mfd}, and also $K$ induced reactions~\cite{Kim:2022bwb, KEK-PSE224:1998trj}.
To estimate the impact of parameter tuning, we use the default parameters of the PACIAE, and find the yield is about 75\% of those obtained with the tuned parameters; the yield of $\Xi$ with the default parameters is about 45\% and that of $N$ is about 170\% of those obtained with the tuned values, as shown in Table \ref{tab: Total yields of baryons}.  
The yields tend to approach zero as  $E_B \to 0$. This is expected because when the binding energy is small, the wave function extends far away, which leads to a vanishing constant $A$ and therefore  the yield goes to zero. 

\begin{figure}[htpb]
\centering
\includegraphics[width=0.45\textwidth]{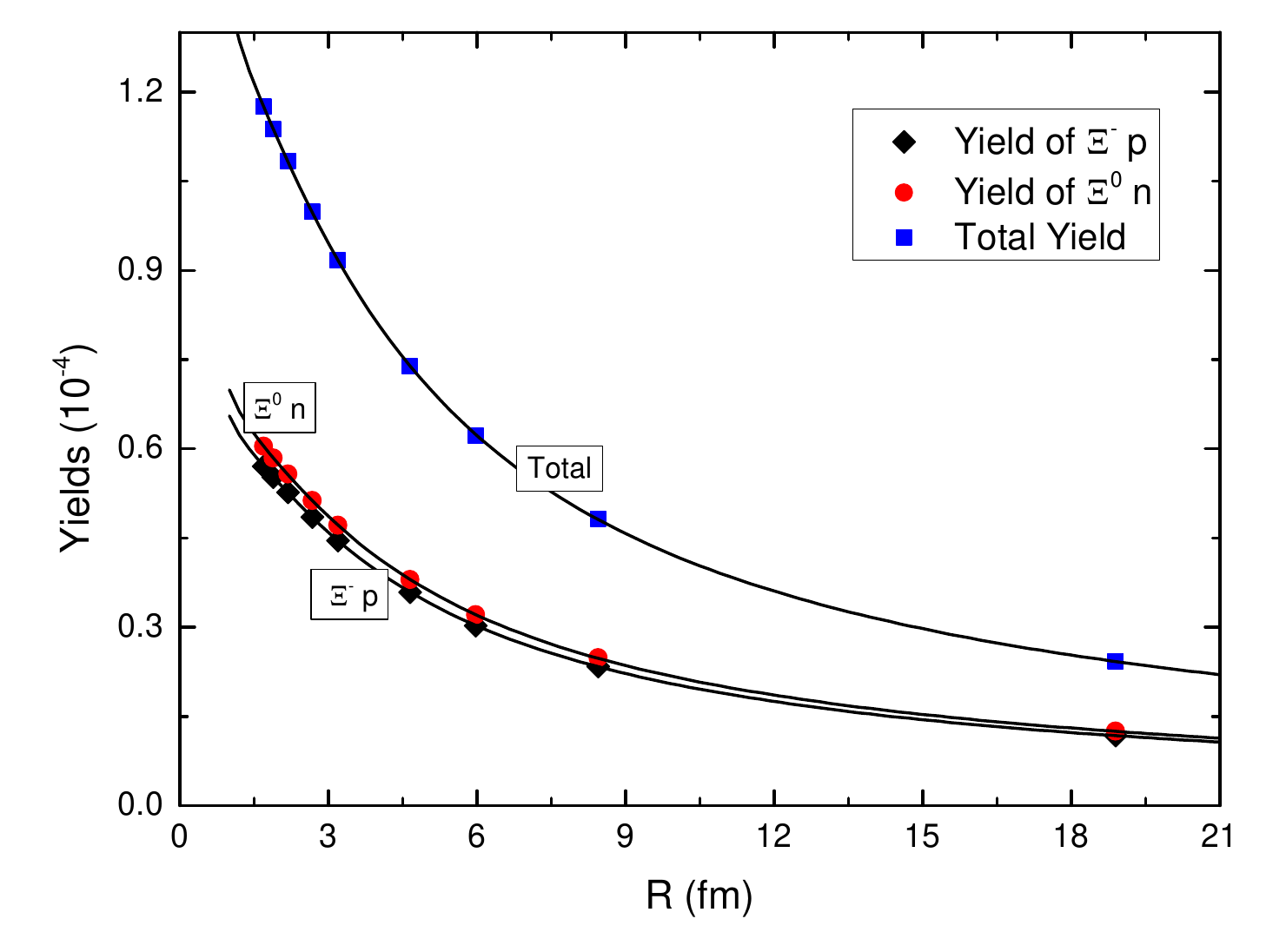}
\caption{ \label{fig: Fitting simulation R} Yield of the $\Xi N$ dibaryon for different root-mean-square radius $R$, where the solid daimonds/squares/circles/ are simulation data, and the lines are analytic results fitted to the simulation data.}
\end{figure}

\begin{figure*}[htbp]
  \subfigure[]{
    \begin{minipage}[t]{0.32\textwidth}
      \centering
      \includegraphics[width=\textwidth]{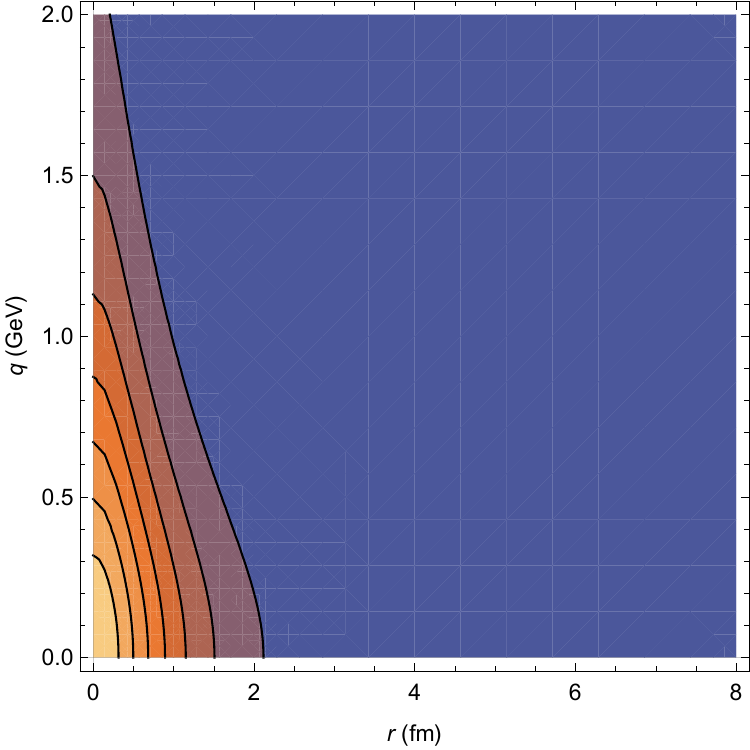}
      \label{fig:density1}
    \end{minipage}
  }
  \hfill
  \subfigure[]{
    \begin{minipage}[t]{0.32\textwidth}
      \centering
      \includegraphics[width=\textwidth]{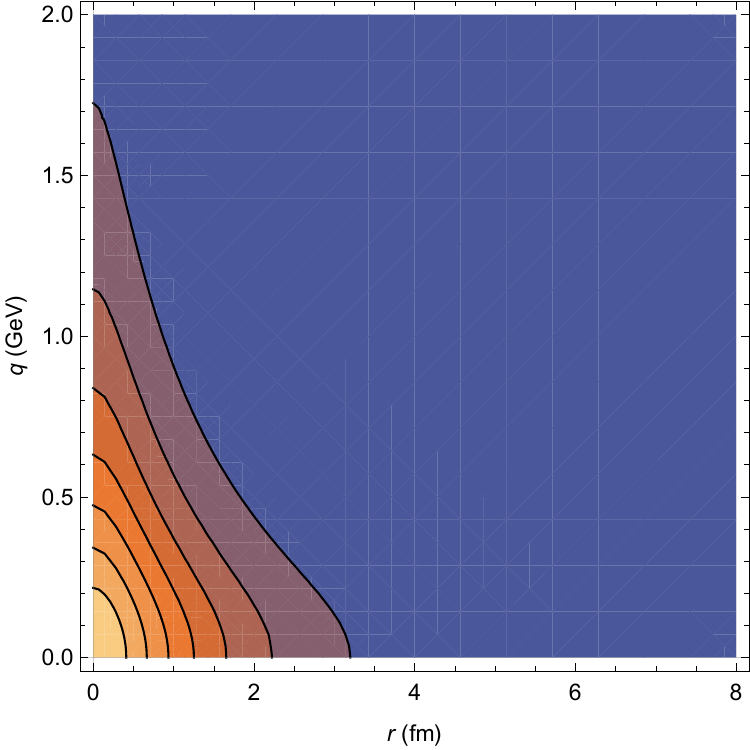}
      \label{fig:density2}
    \end{minipage}
  }
  \hfill
  \subfigure[]{
    \begin{minipage}[t]{0.32\textwidth}
      \centering
      \includegraphics[width=\textwidth]{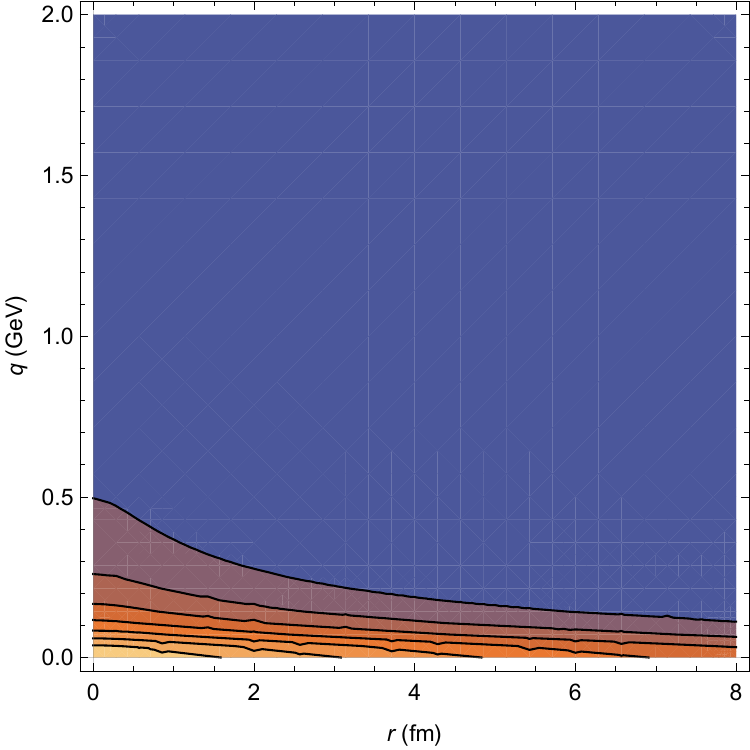}
      \label{fig:density3}
    \end{minipage}
  }
  \caption{\label{fig: Wigner densities of three different size of XiN}The Wigner densities (with the angle between $\bm r$ and $\bm q$ is $\pi/2$) of the $\Xi N$ dibaryons for different binding energies and sizes, where Fig.~\ref{fig:density1}, Fig.~\ref{fig:density2}, and Fig.~\ref{fig:density3} correspond to $E_B=12.5,5$, and $0.1$ MeV, respectively.}
\end{figure*}

\begin{figure*}[htbp]
  \subfigure[]{
    \begin{minipage}[t]{0.32\textwidth}
      \centering
      \includegraphics[width=\textwidth]{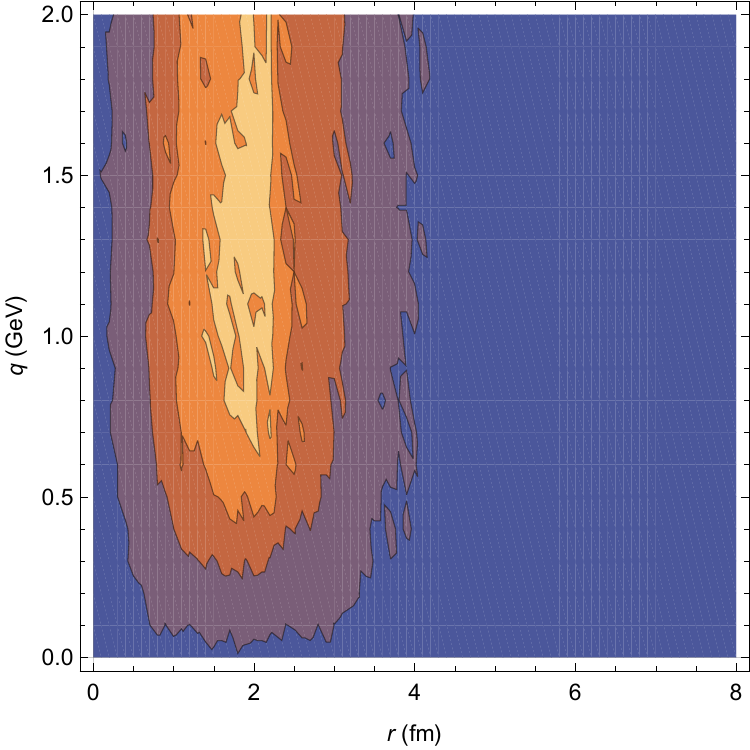}
      \label{fig:phasespace pp}
    \end{minipage}
  }
  \subfigure[]{
    \begin{minipage}[t]{0.32\textwidth}
      \centering
      \includegraphics[width=\textwidth]{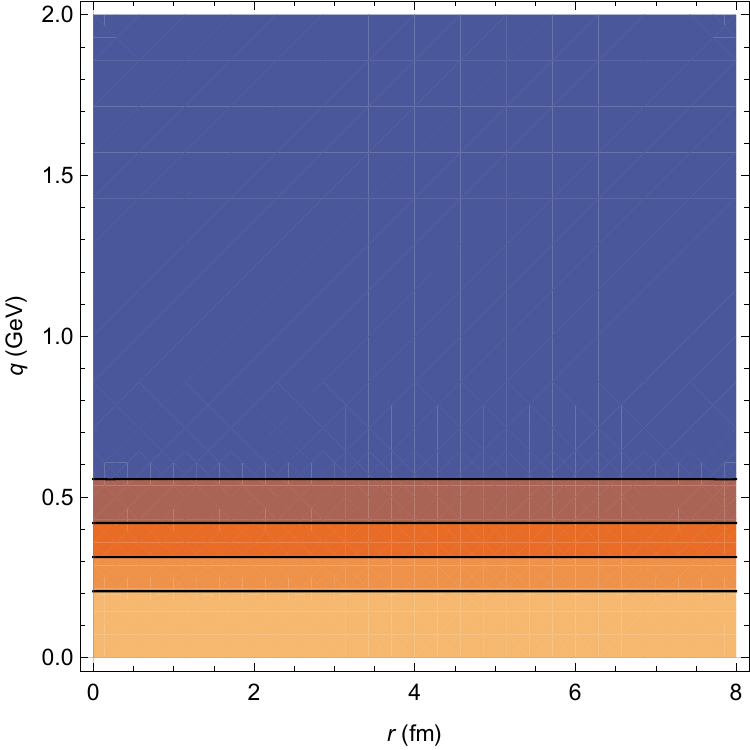}
      \label{fig: phase space heavy-ion}
    \end{minipage}
  }
  \caption{\label{fig: Wigner densities of phase space}Fig.~\ref{fig:phasespace pp} and Fig.~\ref{fig: phase space heavy-ion} are the Wigner densities of the source in the final state phase space of $pp$ collisions and relativistic heavy-ion collisions respectively.}
\end{figure*}

Now we fit the simulation results with the formula of Eq. (\ref{Eq: approximation 2}). The fitted parameters are listed in Table \ref{tab: Fitting parameters}. As shown in Fig. 3, the fits reproduce the simulation data very well with the $\sqrt{E_B}$ dependence for small $E_B$, which is the universality pointed out above.
We have three sets of parameters for the three lines, since $C_1$, $C_2$, and $C_3$ are related to the properties of different sources depending on the configurations. 

We can also interpret the trend in terms of the root-mean-square radius of the system. As shown in Fig. \ref{fig: Fitting simulation R}, the yield of $\Xi N$ is smaller when it has a larger size. This behavior seems contradictory to the one drawn in Refs. ~\cite{ExHIC:2010gcb, ExHIC:2011say}, which claimed that the yield would be larger if the hadronic molecule is more loosely bound. To understand this apparent discrepancy, recall that the yield reflects the overlap of the Wigner density of the source and the composite particle. The Wigner density of the composite particle will extend in the position space and shrink in the momentum space as the size of the composite particle grows, as shown in Fig.~\ref{fig: Wigner densities of three different size of XiN}. However, we note that the Wigner density of the source differs significantly between  $pp$ collisions and heavy-ion collisions as shown in Fig.~\ref{fig: Wigner densities of phase space}. In relativistic heavy-ion collisions, the spatial part is uniformly distributed in phase space since the volume of QGP is very large compared with the size of the composite particle~\cite{ExHIC:2010gcb, ExHIC:2011say}, as the brightest band in Fig.~\ref{fig: phase space heavy-ion} shows. On the contrary, in the final state of $pp$ collisions, the hadrons are produced mainly in the area $r < 4$ fm, centering around 2 fm, as the brightest part in Fig.~\ref{fig:phasespace pp} shows. Intuitively one expects that in $pp$ collisions the overlap will get smaller when the size of the composite particle increases. On the other hand, in the case of heavy-ion collisions, the overlap will get larger since the distribution of the Wigner density is flat. Therefore the phase space distribution in the final state of different collisions is quite different and therefore affects the production yields of composite particles in a nontrivial way. This feature can be used to test the molecular picture of the many exotic hadrons discovered in recent years. 

\begin{table}[htpb]
\caption{\label{tab: Fitting parameters}Fitted parameters for the yields of the $\Xi N$ dibaryon.}
\setlength{\tabcolsep}{3.2pt}
\begin{tabular}{cccc}
\hline
\hline
Parameter & $C_1 (\rm fm \cdot 10^{-3})$ & $C_2 (\rm fm)$&$C_3 (\rm fm \cdot 10^{-3})$\\ 
\hline
$\Xi^- p$ & 0.7957 & 2.3072&0.1828\\
$\Xi^0 n$ & 0.8411 & 2.3531&0.1997\\
Total & 1.6367 & 2.3308&0.3826\\
\hline
\hline
\end{tabular}
\end{table}

\section{Summary}{\label{Summary}}
In this work, adopting the transport model combined with the coalescence model and using the Yukawa-type wave function, we calculated the production yields of the $\Xi N$ dibaryon for different binding energies in $pp$ collisions. For a  binding energy $E_B$ in the range of 0.1 MeV to 12.5 MeV, the yields are about $10^{-4}$, at the same order of magnitude as that of $\Omega$, which indicates that it is possible to discover the $\Xi N$ dibaryon in LHC collisions if it indeed exists and the experimental setup is well-designed.

Furthermore, the yield of $\Xi N$ can be well determined by its binding energy, which is an universal phenomenon. In the small $E_B$ limit, the yield depends linearly on $\sqrt{E_B}$, and  goes to zero as $E_B$ goes to zero. The parameters in the relation encode the phase space information of constituent particles in the kinetic freeze-out stage of $pp$ collisions. We should stress that this phenomenon is only valid for $pp$ collisions since the phase space distributions of the final states in different collision systems play an important role in determining the yield. 

\bigskip
\noindent
\begin{center}
	
	{\bf ACKNOWLEDGEMENTS}\\
	
\end{center}
 We thank Yuyan Xu for valuable discussions. This work is partly supported by the National Natural Science Foundation of China under Grant  No.11975041 and No.11961141004.  T.W. acknowledges the support from the Chinese Scholarship Council. A.H. is partly supported by the Japanese Grant-in-Aid for Scientific Research, Nos. 21H04478 and 18H05407.

\bibliographystyle{elsarticle-num}
\bibliography{BB.bib}

\end{document}